# Heavy-hole band splitting observed in mobility spectrum of p-type InAs grown on GaAs substrate

Jarosław Wróbel*[a], Gilberto A. Umana-Membreno[b], Jacek Boguski[a], Dariusz Sztenkiel[c], Paweł Piotr Michałowski[d], Piotr Martyniuk[a], Lorenzo Faraone[b], Jerzy Wróbel[c] and Antoni Rogalski[a]

a) *Institute of Applied Physics, Military University of Technology, 2 Kaliskiego Str., 00-908 Warsaw, Poland*

b) *School of Electrical, Electronic & Computer Engineering, The University of Western Australia, 35 Stirling Highway, Crawley WA 6009, Australia*

c) *Institute of Physics of Polish Academy of Science, Aleja Lotników 32/46, 02-668 Warsaw, Poland*

d) *Institute of Electronic Materials Technology, Wólczyńska 133, 01-919, Warsaw, Poland*

* email: jaroslaw.wrobel@wat.edu.pl

**Abstract**

*High-quality Be-doped InAs layer grown by MBE on GaAs substrate has been examined via magnetotransport measurements and high-resolution quantitative mobility spectrum analysis (HR-QMSA) in the range of 5 – 300 K and up to 15 T magnetic field. The layer homogenity and dopant concentration has been proofed via HR-SIMS. The results shew 4-channel conductivity and essential splitting of the most populated hole-like channel below 55 K. The multilayer model concluded from the QMSA results has been compared with nextnano simulation.*

## 1. Introduction

The $A^{III}B^{V}$ binary compounds with approximately 6.1 Å lattice constant are currently under the great interest of both research groups [1] and manufacturers of advanced electronics [2]. There are many reasons for the 6.1 Å family popularity. The most important is the extraordinary flexibility for band gap engineering achieved in the infrared energy range via compound alloying and the reduction of dimensionality [3]. Therefore, a wide range of high-speed and low-power devices can be realized, taking the advantage of GaAs, InSb and InAs properties [4]. In particular, a special attention is devoted to ultra-thin layers and low dimensional objects made of InAs, for example quantum dots [5], nanowires [6], topological insulators [7] and quantum wells [8]. Moreover, indium arsenide forms an active layer in various opto-electronic devices, like quantum cascade lasers [9] and infrared detectors [10].

As a result, more and more detailed knowledge about this compound and its interfaces is still needed for the design of complex structures meant for applications.

Especially, the precise description of electrical carrier transport in multi-layered structures is a key factor in the device fabrication. Here, the great progress has been made, in which the possibility of separated-conductivity-channels observation plays an important role. The separation of different contributions to the electrical conductivity has become possible through a technique known as the *mobility spectrum analysis* (MSA) [11]. This method is especially interesting for the characterization of valence bands (VBs) since the interpretation of measurement data obtained for p-type channels is much more demanding than for data collected for transport via normally-separated conduction band. Therefore, there is a relatively small number of literature reports, where multi-channel p-conduction is analyzed and e.g. light-hole effective-mass $\boldsymbol{m_{lh}}$ or hole mobility $\boldsymbol{\mu_h}$ versus temperature are presented [12, 13].

In this paper we present the detailed study on electronic carrier transport in highly doped p-type InAs layer grown on the GaAs semi-insulating substrate. Structure uniformity has been analyzed using the Secondary Ion Mass Specrometry (SIMS) and highly symmetrical 4-terminal electrical devices have been fabricated using the photo-lithography (PL) technique. Resistance and Hall effect have been measured in the temperature range 1.5 to 300 K at magnetic fields up to 16 T. The high quality of obtained data enabled us to apply the very precise variant of MSA method, the so-called High-Resolution Quantitative-Mobility-Spectrum Analysis (HR-QMSA) [11]. We identified up to four distinct conductance channels at low temperatures, one related to electrons, second one to light holes and two to the metallic transport in the heavy-hole band. The last observation has been related to the gradual reduction of strain caused by the lattice mismatch on the InAs/GaAs interface. This interpretation has been supported by the comparison of experimental data with the numerical simulations performed for different temperatures, using the Nextnano code.

## 2. Device fabrication and measurements

The 2 μm p-type InAs epilayer was grown on (001)-oriented semi-insulating epi-ready GaAs substrate via RIBER Compact 21-DZ solid-source molecular beam epitaxy (MBE) system. The layer was deposited on 250 nm-thick GaAs smoothing buffer and high quality of the growth with minimal residual strain has been indicated by XRD and Raman spectroscopy analysis. Details of the growth procedure and basic characterization have been published elsewhere [14]. The InAs layer studied here was doped with Be-dopant of concentration ~ $2\times10^{18}$ cm$^{-3}$ and the doping uniformity has been checked via SIMS analysis using CAMECA SC Ultra system, with depth resolution below 1 nm. The Ga, In, C, O and Be intensities were recorded versus measurement time. Knowing the thickness of the InAs layer from HR-SEM results [14] we determined the sputter rate and depth profile. The results are presented in the Fig. 1a, in which the intensity signal is converted into the volume concentration as indicated on the right-hand axis.

Data shows that the sample is uniformly doped with beryllium. Be, as a light element, can be mixed into the sample during the ion bombardment more effectively than heavier elements like In. Note that the Be decay-length is quite large. However, it depends strongly on the impact energy, so the observed uniformity of distribution is not a SIMS-related artifact. The concentration of carbon and oxygen atoms inside the layers was found to be below the detection limit ($10^{16}$ atoms\cm$^3$ and $3 \times 10^{16}$ atoms\cm$^3$, respectively). Nevertheless, monitoring of these signals allowed us to localize the buffer layer position. Obtained results confirm the assumed architecture of our sample and proof the high homogeneity of the Be-dopant profile, which is important for the mobility spectrum analysis.

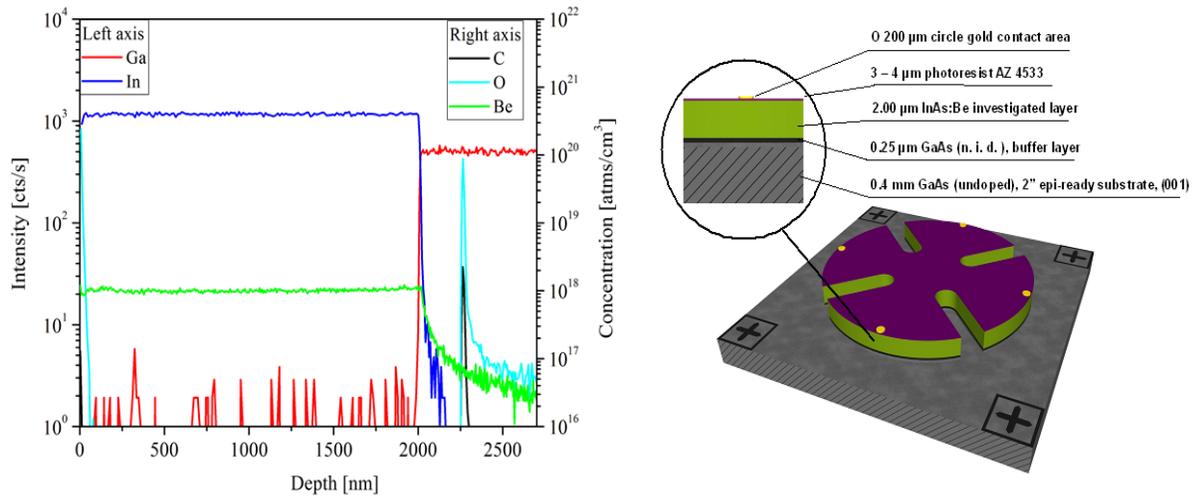

**Fig. 1** (a) SIMS depth profile of the considered layer (left). (b) Schematic view on 5 mm-diameter cloverleaf shape sample and cross section with layer details inset (not in scale, right).

The application of the HR-QMSA method requires a very precise Hall effect measurements, therefore the special attention has been devoted to the fabrication of highly symmetrical samples in the van der Pauw geometry [15], see Fig. 1b. The 4-terminal devices have been defined lithographically using AZ 4533 photoresist and NaOH water solution as a developer. The gold contacts were made by electrolytic deposition using the 1.2 % KAu(CN)$_2$ water solution under 350 – 375 µA/cm$^2$ current density. In the following step, the target cloverleaf shape has been obtained in wet etching process. The surface outside the covered shape was etched down to the GaAs substrate using the orthophosphoric acid: citric acid: hydrogen peroxide: water (molar ratio: 1:1:4:16) solution and 0.4 mol/dm$^3$ hydrochloric acid water solution. Finally, gold wires have been bonded to contact areas. The photoresist has not been washed out after the etching stage, to protect the sample against oxidation during storage.

For the mobility spectrum analysis, not only the shape of a sample is important, but also the linearity of electrical response. Therefore, our devices have been electrically pre-examined in 300 K, in order to check the symmetry and the electrical contact quality by current-voltage linearity test in the voltage range from – 150 mV to + 150 mV. All of the contact pairs gave linear response on voltage bias, where determinacy coefficient for linear function was better than R > 0.999. Using excitation current $I_{exc.}$ = 500 µA, the dissipated power was lower than 0.1 mW, as required [16]. Equally important, is the homogeneity of magnetic field. Therefore, for the purposes of this study the superconducting Cryogen Free Magnet System (CFMS) has been used in which magnetic-field homogeneity ensures ≤ 0.1% total variation over 10 mm sphere diameter [17].

In this system, the Hall sample is placed on the special holder inside the variable temperature insert (VTI), directly in the circulating high-purity helium, which is kept at a constant pressure. Such environment ensures very good thermal conductivity and in consequence the required temperature stabilization ($2\sigma$ in the order of 6 – 20 mK, for 5 – 300 K), monitored by CERNOX sensor. Our measurements have been performed over the ±15 T range for each temperature in the so-called *Step Scan* mode. In order to assure the proper sample temperature, before each magnetic field scan sequence, at least 20 minutes long stabilization have been applied. The ΔB steps have been chosen properly for uniform adjustment into $\log_{10}$ scale, namely 10 points on each decade. The results for two out of twelve resistances are shown in Fig. 2.

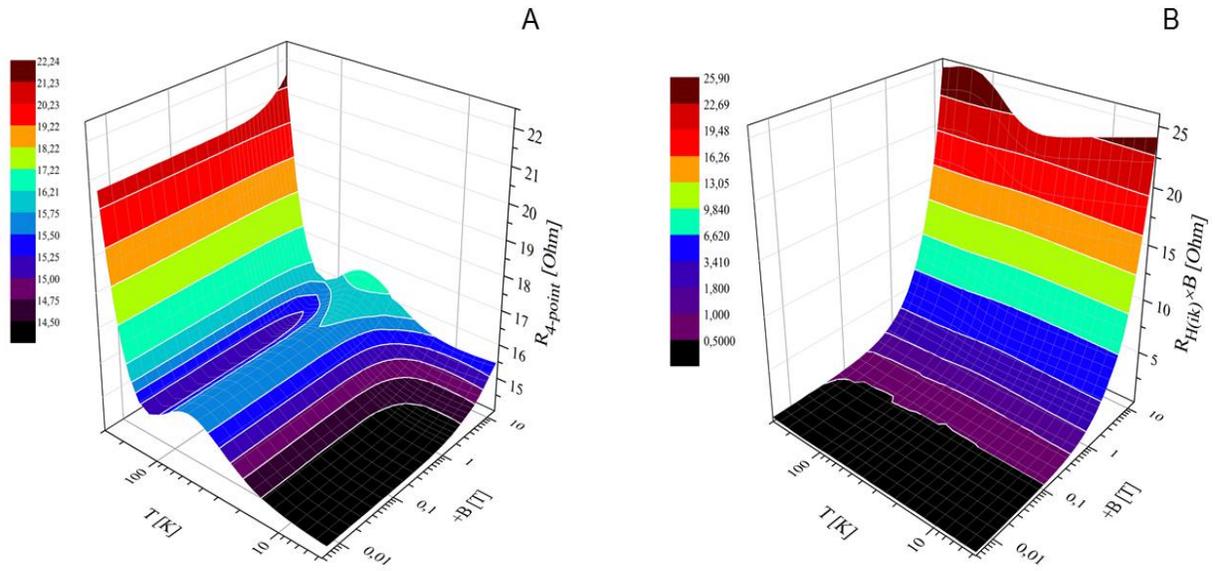

**Fig. 2** (a) The example of experimental van der Pauw 4-point resistance surface plot [$R_{ij} = f$ (T,B)], for one of eight possible configurations. Indices *i* and *j* refer to contact number from figure 1 and each of crossline represent single measurement point. (b) The example of experimental van der Pauw Hall resistance surface plot [$R_{H(ik)}(B) \times B = V_{ji} / I_{ik} = f$ (T,B)], for one of four possible configurations.

As it is seen, the $R_{4-point}$ and the $R_{H(ik)}(B) \times B$ curve families exhibit a high smoothness in the whole ranges of magnetic field and temperature. Moreover, the corresponding surfaces obtained for the opposite magnetic fields direction (not shown) behaved in the highly symmetrical and anti-symmetrical manner, as required by van der Pauw method. Therefore we conclude, that the sample was electrically homogeneous in agreement with SIMS analysis, discussed earlier. Most importantly, the $R_{4-point}$ resistance is clearly non-monotonic as a function of magnetic field and Hall resistances reveal strong non-linearities. It is characteristic for the multi-carrier transport in semiconductors and can be analyzed by the mobility spectrum method.

## 3. Results and discussion

The main concept of mobility spectrum relies on transformation of measurement data from the magnetic field (**B**) domain into the mobility (**μ**) domain. The general idea of such calculations is to obtain the conductivity tensor components $\sigma_{xx}(B)$ and $\sigma_{xy}(B)$ using experimental Hall constant $R_H(B)$ and sheet resistance $R_s(B)$ values, according to the following relations [11]:

$$\sigma_{xx}(B) = \frac{R_s(B)}{R_s^2(B) + R_H^2(B)B^2}, \quad \sigma_{xy}(B) = \frac{R_H(B)B}{R_s^2(B) + R_H^2(B)B^2}.$$

These coupled expressions contain information about all carriers present in the sample according to the discrete mobility transform equations [18]:

$$\sigma_{xx}(B) = \sum_j \frac{S_p(\mu_j) + S_n(\mu_j)}{1 + \mu_j^2 B^2}, \quad \sigma_{xy}(B) = \sum_j \frac{[S_p(\mu_j) - S_n(\mu_j)]\mu_j}{1 + \mu_j^2 B^2},$$

where $S_p(\mu) = e\mu p(\mu)$ and $S_n(\mu) = e\mu n(\mu)$ are the hole and electron *mobility spectra* (i.e., hole and electron conductivities in the mobility domain). The symbols $p_s(\mu)$ and $n_s(\mu)$ are hole and electron sheet densities respectively, e is the electronic charge. Therefore, to identify each type of the multiple carriers contributing to transport we have to obtain mobility distributions $S_p(\mu_j)$ or $S_n(\mu_j)$, using the appropriate numerical methods.

The basic algorithms for that purpose have been described by Beck and Anderson [18], however, they might be insufficient for a proper recognition of multi-channel conductivity in semiconductor devices [19]. Fortunately, a significant progress has been made in recent years and new, improved methods have been successfully tested on many different material systems [20, 21]. Here we applied high-resolution quantitative mobility spectrum analysis (HR-QMSA) [11], which previously gave excellent results for proper recognition of carrier transport trough valence bands in non-intentionally doped GaSb substrate [22]. Results of HR-QMSA method applied to our data are shown in Fig. 3.

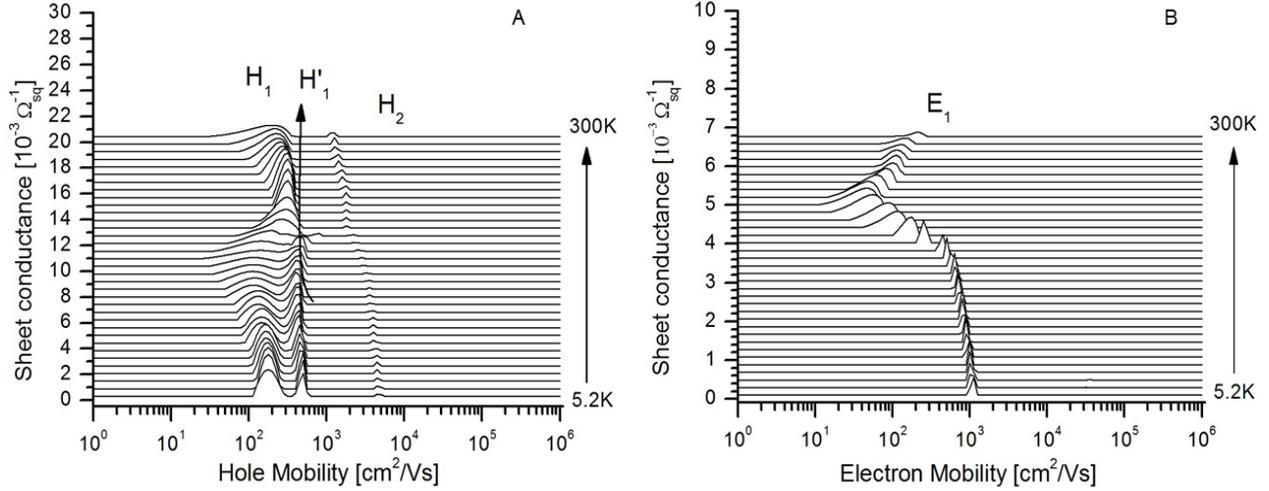

**Fig. 3** (a) Hole mobility spectra $S_p(\mu_j)$ versus temperature Three different hole-like conductivity channels ($H_1$, $H_1{'}$ and $H_2$) have been recognized in low temperature regime. The increase of $\Delta T$ is non-linear. It follows constant step in $\log_{10}T$ scale. (b) Electron mobility spectra $S_n(\mu_j)$ versus temperature. Here, only single conductivity channel is visible. The influence of this channel on total conductivity is less than 3,5 % for all temperatures below 30 K.

Clearly, we identify the four distinct mobility spectra peaks, which contribute to the total conductivity tensor. One is electron-like (denoted $E_1$) and three others are hole-like (denoted $H_1$, $H_1{'}$ and $H_2$). Note that $H_1{'}$ peak disappears for higher temperatures $T > 55$ K or rather merges with stronger $H_1$ contribution. Note also, that $H_2$ feature, which corresponds to carriers with the highest mobility, is rather weak. From the obtained mobility spectra, we calculated the partial conductivity $\sigma_i$, mean Hall mobility $\mu_i^H$ and sheet carrier concentration $N_i$ associated with the $i$-th conductivity peak, using the following formulas

$$\sigma_j = \sum_i S_i(\mu_i), \quad \mu_j^H = \frac{1}{\sigma_i}\sum_i \mu_i S_p(\mu_i), \quad N_j = \frac{\sigma_j}{\mu_j^H}.$$

Results, as a function of temperature, are shown in Fig. 4. The densities $N_j$ and corresponding mobilities $\mu_j^H$ are collected on the left side, the stack area plot of partial conductivities $\sigma_j$ is presented on the right.

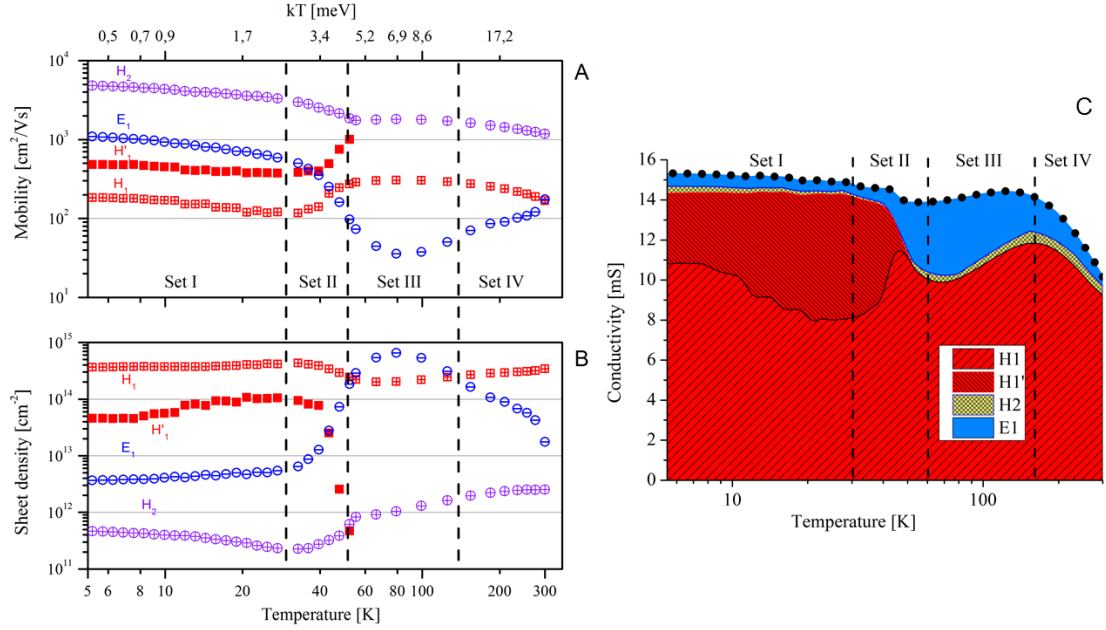

**Fig. 4**. - Hall mobilities (A) and partial densities (B) vs temperature in the log-log scale. The partial contributions to the total conductivity are shown on the right (C) as the stacked area plot. The whole temperature range is divided into characteristic regions, numbered from I to IV.

Both plots reveal at least four characteristic temperature ranges. At $T < 30$ K (region I) and for $T > 150$ K (region IV) all conductivities, in spite of small fluctuations, decrease monotonically and the mobility spectra peaks become wider with increasing $T$. This is not the case for the intermediate range of temperatures (regions II and III), in which the partial conductivities fluctuate very strongly. At the same time, the width of dominant $H_1$, contribution decreases, somehow "at the cost" of electron-like peak $E_1$, which becomes dominant at $T = 80$ K.

We believe, that $E_1$ peak, detected at low temperatures, is associated with the presence of a surface inversion layer, observed on p-type InAs samples since the 1970s [23]. Later, the direct evidence for the existence of charge accumulation on free and clean InAs surfaces were reported, independently on type and doping level [24, 25]. Currently it is widely believed, that InAs has naturally high concentration of surface electrons, which might be e.g. the main source of anomalous effects in temperature behavior of Seebeck coefficient [26]. For example, Olsson et al [25] estimated electron surface concentration on about $10^{12}$ cm$^{-2}$, which compares well with our estimation of $n^{Set\,I} \approx 3 \times 10^{12}$ cm$^{-2}$, at the lowest temperatures. Similar results have been obtained at $T = 300$ K by Lin et al [27] by QMSA method for the surface electrons on n-type sample, which also supports our identification of $E_1$ peak. It is very unlikely, however, that the surface electron concentration reaches the extremely high value of $6.55 \times 10^{14}$ cm$^{-2}$, at $T = 80$ K, as can be observed in Fig. 4. Therefore, the unexpected behavior of $E_1$ peak at the temperature ranges II and III might be an artifact of QMSA method, as we showed by the model calculations, described below.

Clearly, such artifacts related to electrons may also affect the identification of a hole-like mobility spectra. Fortunately, it is not the case for $T < 30$ K and $T > 200$ K, therefore in the following we concentrate on the results obtained at low (region I) and high (region IV) temperatures. Judging from the partial concentrations and mobilities, $H_1$ spectrum is associated with heavy holes, whereas $H_2$ peaks originate from the presence of light holes, in agreement with the recent results of Casias *et al* [28] for very similar 2-μm-thick InAs$_{0.91}$Sb$_{0.09}$ layer, acceptor doped at $3 \times 10^{18}$ cm$^{-3}$ obtained at $T = 300$ K. However, it is not clear what is the origin of the strong splitting of heavy hole spectra, observed at low temperatures, which gradually disappears when peaks broaden and start to overlap.

In order to answer this question, we firstly assumed that the presence of the additional $H_1$ peak is not related to the inhomogeneity of our sample. Indeed, an anisotropy of conducting material can lead to a wrong number of carries species obtained by the mobility spectrum analysis, as it was shown for two-carrier transport in a Hall-bar device [29]. However, the SIMS results (see Fig. 1a) show the excellent uniformity of both parent atoms and Be-dopant. Moreover, we believe that the application of van der Pauv method effectively averages any residual inhomogeneities of our sample. Secondly, we considered the anisotropy (so-called warping) of valence band, which is characteristic for all A$^{III}$B$^V$ materials. We calculated the effective mass tensor for heavy-holes using the InAs parameters which showed that for a spherical Fermi surface and degenerate statistics $\sigma_{xx}$ indeed deviates from the Drude's formula. However, in reality, Fermi momentum $k_F$ depends on crystallographic direction so it should be taken into account in further studies. Anyhow, we assumed that the additional $H_1{'}$ peak *is not* related to QMSA artifact, but to a hole conduction in the degenerate valence band.

Metallic transport and degenerate statistics is suggested by a very weak temperature dependence of partial conductivities, observed up to $T \sim 10$ K. For higher temperatures, the $H_1{'}$ contribution increases, which may suggest hopping-like conduction, which is rather typical for an amorphous films [30]. In our case, the $H_1{'}$ conductivity does not follow Mott' or Efros - Shklovskii' laws [31] and the total contribution $H_1 + H_1{'}$ is practically temperature independent up to 30 K (region I). The above conclusions led us to propose an alternative explanation of the splitting. We inferred that such effect is a result of strained InAs interlayer presence between GaAs substrate and fully relaxed InAs layer. This assumption has been verified via numerical calculations performed in the *nextnano* software [32].

## 4. Model and comparison with experiment

The assumptions about geometry do not differ much from the presented in the figure 1b, except additional 20 nm surface cap layer and 200 nm strained inter-layer. The details have been presented in the figure 5. Beginning from the top part, we have 20 nm surface states that correspond to E$_1$ channel. Unfortunately, proper modelling of the surface effects (especially for InAs) is a difficult challenge. In order to avoid this inconvenience we proposed two numerical treatments. First we use higher than "standard" effective density-of-states electron mass ($m^*_{e_{DOS}} = 0.25$ m$_0$), which corresponds in our model to electron state from asymmetric-near-surface quantum well. Second, we added two extra donor states in order to obtain the effect of electron concentration increasing above 30 K. Their concentrations have been: $n_d = 3.0 \times 10^{19}$ cm$^{-3}$ and $n_d = 5.0 \times 10^{20}$ cm$^{-3}$, with activation energies: $E_{D1} = -15$ meV and $E_{D2} = -5$ meV respectively. Numerical procedures for obtaining desired near surface concentration, like increasing the $m^*_{e_{DOS}}$ value, is rather typical treatment. In this case however more detailed calculations would be needed.

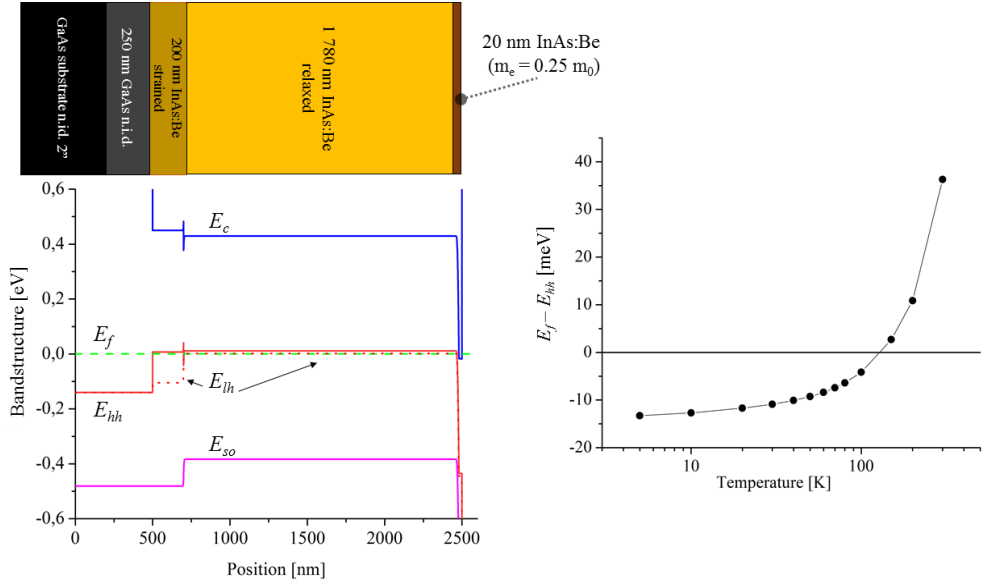

**Fig. 5.** Calculated bandstructure depth profile of assumed geometrical model of InAs layer on GaAs substrate Presented example is for 20 K (left side). The Fermi level position calculated from the InAs midpoint crossing the top of the heavy holes band above 100 K (right side).

Due to expected strain effects, we divided InAs layer (except surface) into two regions. We assumed that the first 1780 nm layer is practically relaxed, with $\varepsilon_{xx} = \varepsilon_{||} = -0.8 \times 10^{-3}$ and $\varepsilon_{zz} = \varepsilon_{\perp} = 0.9 \times 10^{-3}$. It is similar to averaged XRD and Raman measurements results (from this growth) published earlier, in which the parallel and perpendicular residual strain have been determined to be – 1.17 × 10$^{-3}$, and 1.12 × 10$^{-3}$, respectively [14]. The second 200 nm layer, close to the GaAs interface, has been assumed as relatively heavy stained with the parallel and perpendicular residual strain equal to $\varepsilon_{xx} = \varepsilon_{||} = -0.0181$ and $\varepsilon_{zz} = \varepsilon_{\perp} = 0.0197$ respectively. For the both layers the concentration of Be acceptors was $n_A = 3.2 \times 10^{18}$ cm$^{-3}$, with different acceptor ionization energies equal to $E_a = -14$ meV in the first layer and $E_a = -8$ meV in the second one, counted from the top of the heavy hole band. The effective density-of-states mass of electrons and all valence bands, i.e.: heavy-hole ($m^*_{hh\_DOS}$), light-hole ($m^*_{lh\_DOS}$) and split-off band ($m^*_{so\_DOS}$), have been used the same as suggested in the literature, namely: 0.026 m$_0$, 0.41 m$_0$, 0.026 m$_0$ and 0.14 m$_0$, respectively [13]. It should be noted, that the model did not require any special assumptions about the InAs/GaAs interface. GaAs "epi-ready" substrate before InAs deposition has been capped by the 0.5 µm GaAs layer, which effectively covered leftovers of impurities. Thus we didn't expected additional highly conductive inter-layer.

In the bottom left side of the figure 5, the most interesting parts are placed between 500 and 700 nm and close to the surface region. For the first area valence band splitting on over 100 meV can be observed. In our opinion despite of relatively low thickness, this layer is the most probably source of the mobility spectra splitting (see fig. 3a). The latter region, in our model, is the single electron source region in the sample. We rejected here the intrinsic sheet concentration of electrons, as a reliable explanation of the origin of the $E_1$, because even for 300 K, this channel is negligibly small ($\sim 1\times10^8$ cm$^{-2}$). Moreover, we performed qualitative verification of calculations correctness. Namely, the relative Fermi level position in the middle of unstrained InAs layer versus temperature has been checked in order to confirm occurring of the metal-semiconductor transition (right side of the figure 5). The plot crosses the zero value, which means that the Fermi level crossing the top of the heavy-hole band. This process takes place not exactly for 55 K, as in QMSA results, but for about 105 K. The detailed fitting procedures would be needed here, but qualitatively results generally confirms assumed model.

It is worth noting that negative ionization energies for acceptors have been used intentionally in order to simulate metallic conduction. Namely, $E_a$ = - 14 meV in the first layer and $E_a$ = - 8 meV in the second one, counted from the top of the heavy hole band. These values are different from the literature data of $E_a \approx 20$ meV [28]. However it is well known that as the dopant concentration increases, the dopants start to interact and form an impurity band [33-35]. The increase of the width of this band decreases the ionization energy [33-36]. For sufficiently high dopant concentration the associated broad impurity band and the conduction (valence) band edge merge leading to the negative ionization energies.

In the final step, the comparison between presented *nextnano* simulation and HR-QMSA analysis has been performed. The results have been gathered in the figures 6a and 6b. Except transitional regime and the results for channel $H_1'$, figure 6a shows good agreement between two methods. Howefer the obtained numerical curves have no additional peaks for 30 K < T < 110 K and they behavior are smoother than the experimental ones. In our opinion, this comparision suggests also that unexpected apperance of $E_1$ peak might be an artifact of QMSA method.

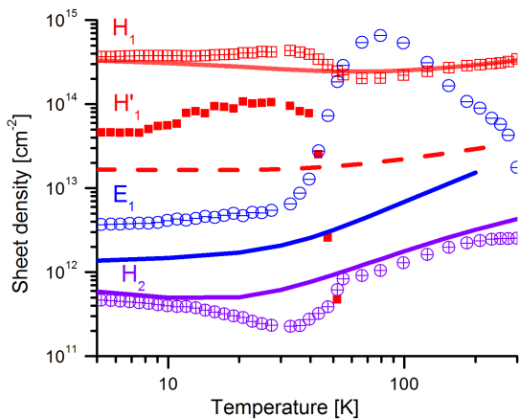
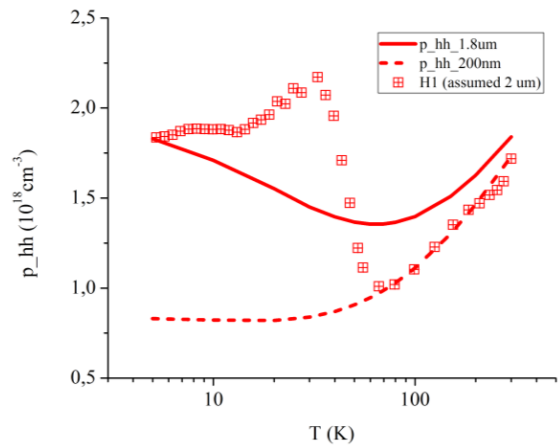

**Fig. 6a.**– Comparison of *nextnano* calculation results (solid lines and dashed line for heavy-hole counted for

**Fig. 6b** – Normalization to the assumed layers thicknesses and changing units into cm$^{-3}$ show

200 nm InAs inter-layer) with sheet densities obtained by Hall measurements and HR-QMSA analysis (colored dots with the same description as in the fig. 4a). that volume concentration of both: relaxed 1.8 um and strained 0.2 um are nearly the same.

In Fig 6a, normalization to the depth profile has not been applied. It has been made in the next step, in which normalization to the assumed layers thicknesses and changing units into cm$^{-3}$ show that volume concentration of both 1.8 um relaxed and 0.2 um strained are nearly the same in the Boltzman statistics regime. We infer here, that the same volume concentrations might be the reason of disappearing of $H_1'$ as a separate channel for $T > 55$ K. In our opinion this is the most probably explanation of the observed heavy-hole band splitting via HR-QMSA method.

## 5. Conclusions

The set of arguments have been presented that observed splitting of the heavy-hole band mobility spectra origin from the thin, strained InAs interlayer between GaAs substrate and fully relaxed InAs layer, despite its very high quality confirmed via multiple characterization techniques. Apart from the heavy-holes we identified also two additional channels.: one related to electrons, second one to light holes in the whole temperature range: 5 -300 K. These results confirm usefulness of the HR-QMSA technique even for the degenerate statistic regime. This interpretation has been supported by the comparison of experimental data with the numerical simulations performed for different temperatures, using the nextnano code.

## Acknowledgments

This study was carried out with the financial support from the Polish National Science Center (grant no. UMO-2015/17/B/ST5/01753).